\documentstyle[preprint,aps]{revtex}

\begin{document}

\thispagestyle{empty}

\title{Non-perturbative Gluons and Pseudoscalar Mesons
in Baryon Spectroscopy}

\author{
Z. Dziembowski$^a$, M. Fabre de la Ripelle$^b$, and
Gerald A. Miller$^c$
\vspace{10pt} }

\address{$^a$ Department of Physics \\ Temple University \\
Philadelphia, PA 19122
\vspace{10pt}}

\address{$^b$
Institut de Physique Nucleaire, \\Universite de Paris-Sud,
94106 Orsay, France\vspace{10pt}}

\address{$^c$ Department of Physics %%FM-15
\\University of
Washington, Box 351560 \\ Seattle, WA 98195-1560 \vspace{15pt}}

\maketitle

\begin{abstract}
We study baryon spectroscopy including  the effects of
pseudoscalar meson exchange and one gluon
exchange potentials between quarks, governed by $\alpha_s$. The
non-perturbative,
hyperspherical method calculations  show that  one can obtain a good
description of the data
by using a quark-meson coupling constant that  is %%
compatible with the
measured pion-nucleon coupling constant, %%
and a reasonably small value of $\alpha_s$.
\end{abstract}

\pacs{12.39pn, 13.75.Gx, 14.20.-c, 21.30.-x}
\newpage

% begin the body
Interest in studying baryon spectroscopy has been re-vitalized by
the recent work of Glozman and
Riska\cite{GLO1,GLO2,GLO3,GLO4,G5}.
These authors point out the
persistent difficulty in obtaining a simultaneous description
of the masses of the  P-wave baryon
resonances and the Roper-nucleon mass difference. In particular
they argue
\cite{GLO2}
that ``the spectra of the nucleons, $\Delta$ resonances and the
strange hyperons are well described by the constituent quark
model, if in addition to the harmonic confinement potential the
quarks are assumed to interact by exchange of the $SU(3)_F$ octet
of pseudoscalar mesons". Furthermore, Ref. \cite{G5}
states that gluon exchange has no relation with the spectrum
of baryons
!

The ideas of Glozman and Riska are
especially interesting because of the good
descriptions of the spectra obtained in
Refs.~\cite{GLO1}-\cite{G5}, and because of the contradictory
long-standing belief
\cite{DERU,IGK1,texts,AE86} that one-gluon exchange is a basic 
element of quantum chromodynamics QCD and the
success of that 
interaction in baryon spectroscopy. 
Despite the lore, %%
 some authors had noted the difficulty
in obtaining a simultaneous description of the Roper and P-wave
resonances\cite{OS82,SB85}.

The purpose of 
this paper is to include both effects in
calculating the baryon spectra using a non-perturbative technique,
 and to show that both kinds of
effects are required for a reasonable description of the data.
Including the effects of pion clouds is known to lead to a good description
of nucleon properties, as well as meson-nucleon and electron-nucleon
scatterings\cite{CBM,ZD}. We note that several previous
 workers\cite{HJW}-
-\cite{FF} have shown that including both
pion exchange and gluon exchange effects leads to an improved
description of the data. Those calculations use a perturbative
treatment of the pion and gluon exchange interactions.

However, non-perturbative calculations are required to handle the
one-gluon exchange interaction\cite{SB85,CI86,KCP}.
It is therefore natural to expect that
if one
used only pseudoscalar meson
exchange to generate all of the mass splitting, a
non-perturbative treatment would be necessary. Thus a
non-perturbative, all-orders treatment is needed to assess
whether or not either of those two elements can be
ignored. We employ the hyperspherical methods of
Fabre de la Ripelle et al \cite{DLF87}
to compute the energies of the baryons.

We use a constituent quark model 
Hamiltonian which includes the effects of one gluon
exchange (OGE) and the exchange of pseudoscalar mesons mandated by
broken chiral symmetry, $V_\chi$, in addition to the kinetic
energy and confinement terms. Thus
\begin{equation}
H=T+V_{con}+V_{OGE}+V_\chi,
\end{equation}
where the kinetic energy $T$ takes the non-relativistic form
\begin{equation}
T=\sum_i -{\nabla_i^2\over 2 m},
\end{equation}
with  the $u$ or $d$ quark mass taken as  336 MeV to represent 
the non-perturbative effects which influence the properties of a 
single confined quark. 
We limit ourselves to light quarks in this first 
calculation, but note that 
the success in handling strange baryons in an important part of the
work of Glozman and Riska.

Here we assume that
the confining interaction
$V_{con}$ takes on a 
linear ($V_L$) form so
that:
\begin{equation}
V_L=\sum_{i< j}A_L\;\mid \vec r_i-\vec r_j\mid.
\end{equation}
The parameter $A_L$ is to be determined
phenomenologically. The one gluon exchange interaction between 
different quarks is given
by the standard expression
\begin{equation}
V_{OGE}=\sum_{i<j}[-{2\over 3}{\alpha_s\over r_{ij}} +{2\over 3}
{\pi\alpha_s\over m^2}{1\over 4\pi}{e^{-r_{ij}/r_0}\over r_0^2\;r_{ij}}\\
-\alpha_s{4\over 9}{\pi\over m^2}{1\over 4\pi}{e^{-r_{ij}/r_0}
\over r_0^2\;r_{ij}}\,
\vec{\sigma}_i\cdot\vec{\sigma}_j],
\end{equation}
where $r_{ij}\equiv \mid\vec r_i-\vec r_j\mid,\;r_0$= 0.238 fm,
and
 $\alpha_s$ is a parameter to be
determined phenomenologically. The replacement of the usual
delta function form by a Yukawa of range $r_0$ is
intended to include the effects of the finite sized nature of
the constituent quarks. 

We ignore the spin-orbit and tensor
terms because 
our first calculation is intended to be a broad comparison of the 
non-perturbative 
effects of gluon and meson exchange.
Isgur and Karl \cite{IKplb} found 
that including 
the tensor hyperfine forces with relative strengths predicted 
by the 
one gluon exchange interaction is necessary to produce the splitting between 
the $J^\pi=1/2^-$ and $J^\pi=3/2^-$  nucleonic states as well 
as to understand their separate wave functions and consequent
decay properties. 
Therefore we do not expect our calculations to reproduce those features.
The issue of the spin-orbit interaction between quarks is a 
complicated one. There are many different contributions: Galilei 
invariant and non-invariant terms arising from one gluon 
exchange see e.g. \cite{v95}, a Thomas precession term arising
 from the confining 
interaction\cite{IGK1}, effects of exchange of scalar mesons
and the instanton induced 
interaction\cite{ST}. The above cited authors show that 
some of the various terms tend to cancel
when evaluating the baryon spectra.
A detailed study of the influence of the various contributions to 
the spin orbit force is beyond the scope of the present work.

The effects of pseudoscalar meson octet exchange are
described by the interaction\cite{GLO1}-\cite{G5}
\begin{equation}
V_\chi=\sum_{i<j}\alpha_{q\pi}{\vec\sigma_i\cdot\vec\sigma_j\over
3}{\vec{\lambda}^F_i\cdot\vec{\lambda}^F_j\over 4
\;m^2}[\mu^2{e^{-\mu\;r_{ij}}\over
r_{ij}}-{e^{-r_{ij}/\Lambda}\over\Lambda^2\;r_{ij}}],
\end{equation}
where $\Lambda=0.238$ fm \cite{AB} represents the
effects of the finite size of the constituent quarks. We shall 
allow the strength of the meson exchange potential,
$\alpha_{q\pi}$,    %%
to vary away from the expected \cite{GLO2} value of 0.67. This
is   in the spirit of the work of  %%
Refs.~\cite{GLO1}-\cite{G5} who fit a very few matrix elements
of  $V_\chi$ to a few mass differences and predict the remainder 
of the
spectrum.
The values of the flavor SU(3) matrices are taken from
Eq. (5.1) of Ref.\cite{GLO2}. We neglect the tensor force 
generated by the exchange of pseudoscalar mesons, as do Glozman 
and Riska. Similarly, retardation effects     
and the influence of the  baryonic mass differences are neglected.

Next we turn to a brief description of the hyperspherical method,
which has been in use for some time\cite{DLF87,APP}. The idea is
that the Schroedinger equation for three particles can be
simplified by expressing the usual Jacobi coordinates
$\vec \xi_1=\vec{r}_1-\vec{r}_2$ and $ \vec{\xi}_2\equiv
{1\over \sqrt{3}}(\vec{r}_1+\vec{r}_2-2\vec {r}_3)$
using the hyperspherical coordinates defined by a radial distance
$r=\sqrt{\xi_1^2+\xi_2^2}$, polar angles
$\omega_i=(\theta_i,\phi_i)$ of $\vec \xi_i$, and the additional
angle $\phi$ defined as tan$\phi=\xi_2/ \xi_1$.
The hyperspherical harmonics consist of a complete set of angular
functions on the 5-dimensional hypersphere. Hence the wave
function and potential can be  expressed in terms of linear combinations
of these functions. Furthermore, Ref.~\cite{Simonov} has shown how to
construct linear combinations of these functions that form
irreducible representations of the permutation group of three
particles
in the S-state. This enables one to construct wave functions that are
consistent with the Pauli exclusion principle. In particular,
the requirement of constructing
color-singlet states is met by treating the wave function as a
product of  the standard SU(6) spin-flavor wave functions, by
symmetric spatial wave functions, by the anti-symmetric color
wave  function.
Therefore
the effects of mixed symmetry states are 
ignored here.

The basis of hyperspherical harmonics has a large degeneracy,
which can be handled by using the optimal subset\cite{fabre}
which is constructed
as linear combinations of Potential Harmonics, i.e.,
those states generated by allowing
 the potential
$V_{con}+V_{OGE}+V_\chi$ to act on the hyperspherical harmonics
of minimal order allowed by the Pauli exclusion principle.
See Ref.\cite{APP} for  a detailed discussion of the general formalism.
The convergence properties of the expansion and the accuracy 
of using a single optimal state have been studied by several authors
\cite{jmr,mns} with the result that the overlap between the 
approximate and exact eigenfunctions is generally greater than 
99.5\%. 

To be specific, we display the specific nucleon and $\Delta$ wave 
functions. The nucleon wave
function is   given by
\begin{equation}
\psi^N=\frac{1}{\sqrt{2}}\,\left[\chi^\rho\eta^\rho+\chi^\lambda\eta^
\lambda\right]
u_N(r)\,r^{-5/2},
\end{equation}
where $\chi^\rho,(\eta^\rho)$ are the mixed antisymmetric spin
(flavor) wave functions and $\chi^\lambda,(\eta^\lambda)$,
are the mixed-symmetric spin (flavor) wave functions. The
$\Delta $ wave function is given by
\begin{equation}
\psi^\Delta=\chi^{3/2}\eta^{3/2} u_\Delta(r)\,r^{-5/2}.
\end{equation}
 The
radial wave functions $u_N$ and $u_\Delta$ are  obtained by solving the
differential equation:
\begin{equation}
\left[{\hbar^2\over m}\left(-{d^2\over dr^2}
+{15/4\over r^2}\right)+V_{N,\Delta}(r)-E\right]u_{N,\Delta}(r)=0,
\label{eqn}\end{equation}
where the potentials $V_{N,\Delta}(r)$ are obtained by re-expressing the
interactions above in terms of  a quark -quark
interaction $V_{qq}$ such that
\begin{equation}
V_{qq}(r_{ij})=V^0(r_{ij})+V^S(r_{ij})\,\vec\sigma_i\cdot\vec\sigma_j
+V^\chi(r_{ij})\,\vec\sigma_i\cdot\vec\sigma_j
\,\vec{\lambda}^F_i\cdot\vec{\lambda}^F_j.
\end{equation}
The term $V^0$ includes both the confining and spin independent part
of the quark-quark interaction.
Then the potential $V(r)$ of Eq. (\ref{eqn}) is given by
\begin{eqnarray}
V_N(r)={48\over \pi}\int^1_0\left[V^0(r\;u)-
V^S(r\;u)+C_N\;V^\chi(r\;u)\right]\sqrt{1-u^2}\;u^2\;du,\\
V_\Delta(r)={48\over \pi}\int^1_0\left[V^0(r\;u)+
V^S(r\;u)+C_\Delta\;V^\chi(r\;u)\right]\sqrt{1-u^2}\;u^2du,\nonumber
\end{eqnarray}
where $C_N= 14/3 $ and $C_\Delta= 4/3$ are
obtained by taking the matrix elements of the flavor-spin matrix
$\vec\sigma_i\cdot\vec\sigma_j\,\vec{\lambda}^F_i\cdot\vec{\lambda}^F_j$ in the
appropriate wave
functions. 
The differential equations are solved using the renormalized Numerov
method formulated by Johnson~\cite{john73}.

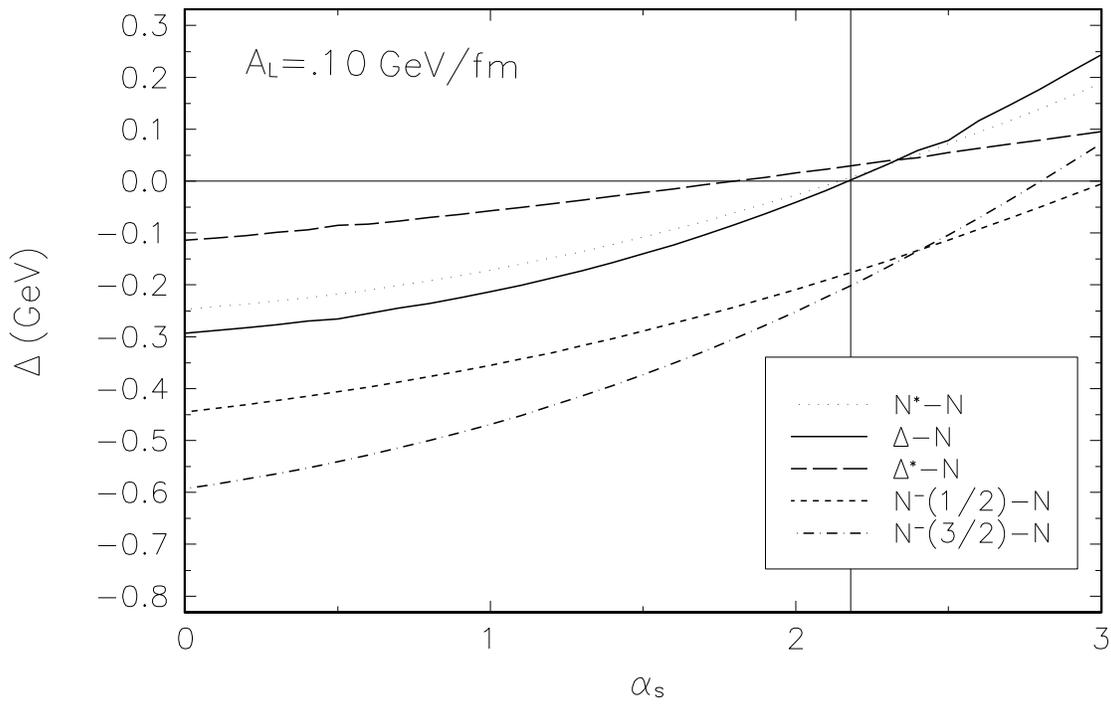
\begin{figure}
\caption{
Baryon mass splitting versus $\alpha_s$, with  $V^\chi=0$.
Differences between the computed
and measured values of the mass splitting $\Delta$ (in GeV) are shown.
a) A=0.10 GeV/fm b) A=0.45 GeV/fm.
}
\label{nopi}
\end{figure}

The first model we shall consider includes the one-gluon exchange
but neglects
the effects of the meson exchange interaction 
$V_\chi$. The
differences between the computed
and measured values of the mass splitting are shown as a
function of $\alpha_s$ in
Fig.\ref{nopi}.
A curve passes through the 
horizontal line when the computed value of the indicated mass difference
is equal to the experimental value of that difference.
This notation is used in each of the figures.
The results of
Fig.\ref{nopi}a show how the model can account for the
splitting between the $\Delta$ and nucleon, $\Delta^*$ and
the $\Delta$,
and the Roper and nucleon, but not the splitting between the
P-wave resonance and the nucleon. Note %%
that a large value of $\alpha_s\approx 2.2$ is used to obtain the fit
with A=0.10 GeV/fm. If one uses instead  A=0.45 GeV/fm, one is able
to account for the $\Delta$-nucleon and P-wave nucleon splitting
but not the Roper,   %%
as shown in Fig.~1b. This agreement is obtained also for a large  value
of $\alpha_s\approx 1.4$ that roughly corresponds to
the original theory  of Refs. \cite{DERU,IGK1} which
works reasonably well except for the Roper.

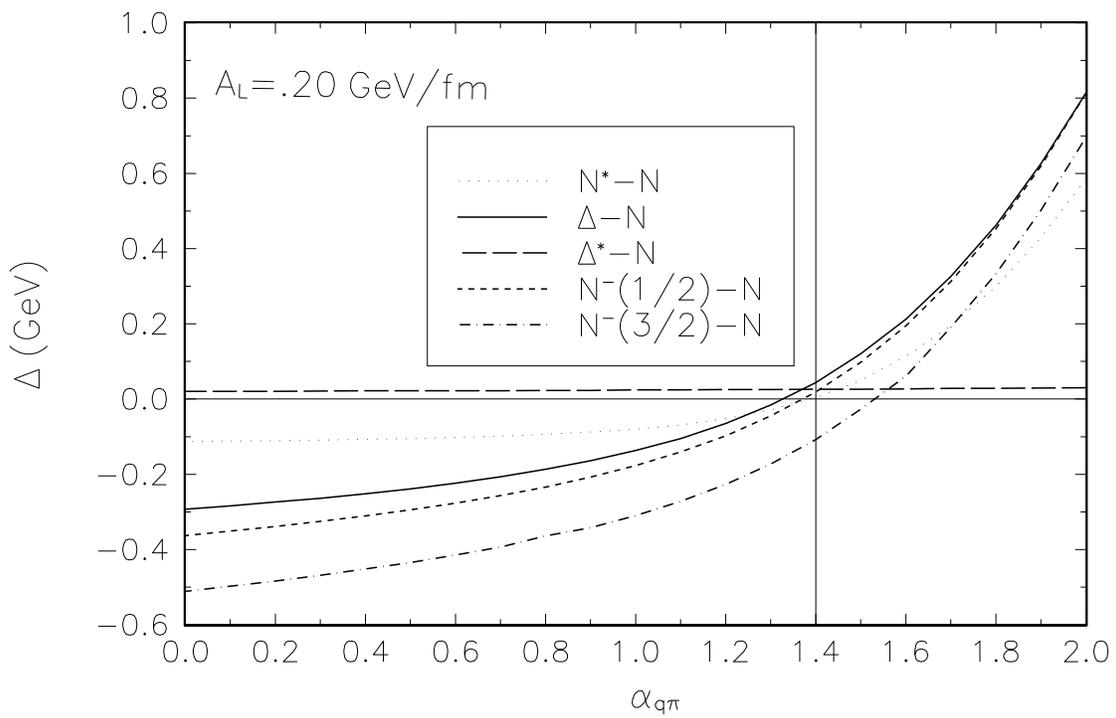
\begin{figure}
\caption{
Baryon splitting- Glozman Riska model,  neglecting the
one gluon exchange interaction, $V_{OGE}=0$.
The mass differences ($\Delta$ in GeV) are
shown as a function of $\alpha_{q\pi}$.
}
\label{noglue}
\end{figure}

One may also study
the converse situation of keeping pseudoscalar meson exchange and
ignoring the gluonic exchange, 
which represents a non-perturbative treatment of the 
Glozman-Riska theory.
 The results, shown in
Fig.~\ref{noglue}, indicate that  this version of the
non-relativistic quark model is very successful if one allows the
freedom to vary the value of $\alpha_{q\pi}$ away from the 
expected value of 0.67\cite{GLO2}.%%
 Using a factor of two increase so that 
$\alpha_{q\pi}\approx 1.4$ improves
immensely the agreement with experiment. No such agreement can be
obtained if one insists on using the value 0.67. 
Note also that the energy of the 3/2$^-$ state is not too well
described.

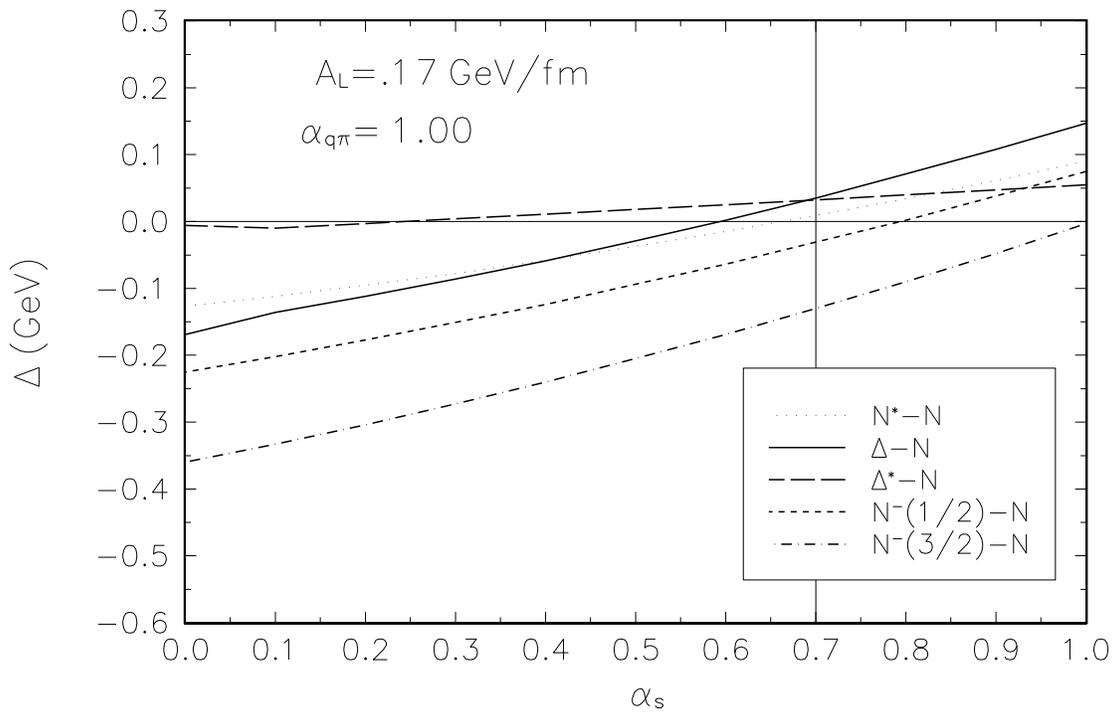
\begin{figure}
\caption{
Baryon splitting with the complete Hamiltonian.
The mass differences ($\Delta$ in GeV) are
shown as a function of $\alpha_s$}
\label{full}
\end{figure}

The third model we consider is the most general, in which both
the color magnetic and pseudoscalar meson exchange terms  are
included.  
Both of these terms contribute to the N-$\Delta$
splitting\cite{CBM}, so that including both effects can be
reasonably expected to lead to smaller values of
$\alpha_s$ and $\alpha_{q\pi}$ than used in Figs. 1 and 2.
The results for this general model are shown in Fig.~\ref{full}. One 
obtains a
good description of the data, with   
the energy of the state N3/2$^-$ state as the expected
single exception.
Furthermore, the value of $\alpha_s$ is about 0.7 instead of
about 2 required if this  is the sole physics responsible for  the
$\Delta$-nucleon mass splitting. A smaller value is preferred
because this interaction is derived using perturbation theory.
Still another  nice feature is that the value of $\alpha_{q\pi}\approx 1
$
which is close the value expected from the measured pion nucleon 
coupling constant, g$_{\pi N}$. The 
relation between the pion-quark coupling constant, g,
and g$_{\pi N}$ is g$_{\pi N}={m_u\over g_A m_N} g$\cite{GLO2}
Using the experimentally measured axial coupling constant 
$g_A=1.26$
along with our quark mass $m_u $= 336 MeV and 
${g^2_{\pi N}\over 4\pi} =14.2 $ gives $\alpha_{q\pi}= {g^2\over 
4\pi} $=1.1. The use of $g_A=1.26$  accounts for known  relativistic 
effects, which change the quark wave functions but do not modify 
the spectrum\cite{texts}.
The use of  $\alpha_{q\pi}\approx 1
$ to reproduce the differences between baryon masses therefore 
represents a significant 
improvement in the theory.

We have obtained
a good description of the energies of states, 
so that it is worthwhile to begin discussing some of the 
properties of the wave functions.
We note that the 
value of $A_L=.17$ GeV/fm, which yields  %%
a nucleon rms radius of 0.46 fm 
is significantly smaller than the experimental value
$\sim$0.8 fm, but much larger than obtained, $\approx 0.3$ fm,
in work using only
one gluon exchange such as that of Refs.\cite{CI86,KCP}. %%
We note that including
the relativistic recoil correction, also
invoked by Capstick and Isgur, is known to increase the computed
value of
the radius. Similar effects occur by including
the influence of the meson cloud on
the nucleon radius, and the effects of other components
of the wave function. We plan to include such effects, along with 
 tensor and  spin orbit forces and retardation effects
in future work. This would enable us to obtain a realistic 
treatment and to compute the decay properties of the excited 
states.
We also plan to %%
consider strange baryons.

The net result of the present work  is that 
non-relativistic
calculations including confinement, one gluon and pseudoscalar
meson exchange can describe the light-quark baryon  spectrum
reasonably well. Most of the mass differences between the states 
are described within accuracy of 10 \% or better.
In particular, we find that including the effects of meson
exchange leads to a good simultaneous description of the Roper-N
and P-wave-nucleon splitting even if the one gluon exchange
interaction is neglected. This is in agreement with Riska
and Glozman. %However, the idea that gluon exchange can be ignored
%all together is shown to be wrong.
However, both gluonic and
pseudoscalar meson exchange are expected from the underlying
theory. One also gets a good description of the baryon
energies in this more general
theory, with the improvements that
the value of $\alpha_s$ is smaller than before  and the value
$\alpha_{q\pi}$ is very  close
 to the one provided by the
pion-nucleon coupling constant.
 Thus although we verify several of the statements of 
Refs.~\cite{GLO1}-\cite{G5}, a theory which includes 
both gluon and meson exchange seems
more plausible.  

We thank E.M. Henley and C.M. Shakin for  useful discussions.
Z.D. thanks the U.W. Physics Education and Nuclear Theory groups for their
hospitality during this work. The work of G.A.M. is supported 
in part by the USDOE.

\newpage

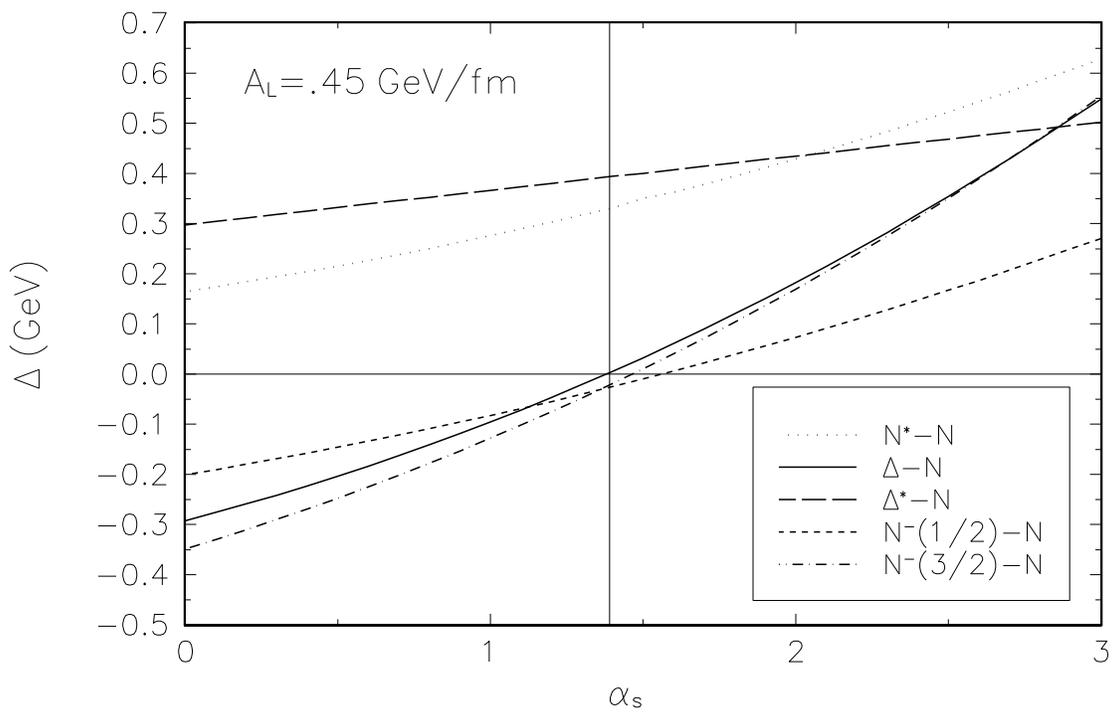
\end{document}